\documentclass[aps,floats]{revtex4}
\usepackage{amsmath,amssymb}
\usepackage{graphicx,epsfig}
\begin{document}
\bibliographystyle {plain}

\def\oppropto{\mathop{\propto}} 
\def\opsimeq{\mathop{\simeq}}
\def\opoverderline{\mathop{\overline}}
\def\operarrow{\mathop{\longrightarrow}}
\def\opsim{\mathop{\sim}}

\def\fig#1#2{\includegraphics[height=#1]{#2}}
\def\figx#1#2{\includegraphics[width=#1]{#2}}

%\newcommand{\fig}[2]{\epsfxsize=#1\epsfbox{#2}} \reversemarginpar 

%%%%%%%%%%%%%%%%%%%%%%%%%%%%%%%%%%%%%%%%%%%%%%%%%%%%%%%%%%%%%%%%%%%%%%%%%%%%
\title{ Random elastic networks : strong disorder renormalization approach } 

%%%%%%%%%%%%%%%%%%%%%%%%%%%%%%%%%%%%%%%%%%%%%%%%%%%%%%%%%%%%%%%%%%%%%%%%%%%%

 \author{ C\'ecile Monthus and Thomas Garel }
  \affiliation{ Institut de Physique Th\'{e}orique, CNRS and CEA Saclay,
 91191 Gif-sur-Yvette, France}

\begin{abstract}
For arbitrary networks of random masses connected by random springs, we define a general strong disorder real-space renormalization (RG) approach that generalizes the procedures introduced previously by Hastings [Phys. Rev. Lett. 90, 148702 (2003)] and by Amir, Oreg and Imry [Phys. Rev. Lett. 105, 070601 (2010)] respectively. The principle is to eliminate iteratively the elementary oscillating mode of highest frequency associated with either a mass or a spring constant. To explain the accuracy of the strong disorder RG rules, we compare with the Aoki RG rules that are exact at fixed frequency.

\end{abstract}

\maketitle

 \section{ Introduction} 

In the field of disordered systems, the problem of random masses
connected by random springs is very old since it has been 
 introduced by Dyson \cite{dyson} even before the classical 
Anderson localization paper \cite{anderson}.
After studies concerning the one-dimensional case
(see the review \cite{ishii} and references therein), 
an analysis of disordered elastic media
via a non-linear sigma-model \cite{john_Sompo} has predicted results similar to 
the scaling theory of Anderson localization \cite{scaltheo} :
all finite-frequency phonons are localized in dimension $d \leq 2$, whereas there exists a finite critical frequency transition between delocalized and localized modes in dimension $d>2$.
More recent discussions on the similarities and differences with 
Anderson localization of electrons 
can be found in \cite{elliott,Leb_spo,us_phonon} and references therein.

Besides the case of regular lattices, one may also consider the case
of more general networks with various physical motivations.
Two types of strong disorder real-space renormalization (RG) procedures
have been previously introduced:
Hastings \cite{hastings} has proposed to eliminate iteratively
masses, whereas Amir, Oreg and Imry \cite{imry}
have proposed to eliminate iteratively springs. 
In these two studies, the results have been found to be extremely accurate
at low energies. In the present paper, we unify these two particular
procedures into a single framework,
 where both masses or springs can be iteratively
eliminated in a consistent way. In addition, we explain the accuracy
of strong disorder RG procedures at low frequency in these models,
by a comparison with the Aoki RG rules that are exact at fixed frequency
(this type of renormalization has been first introduced by Aoki 
in the context of Anderson localization in \cite{aoki80,aoki82,aokibook}).

The paper is organized as follows.
Section \ref{models} presents the models and notations.
In section \ref{sec_strong}, we describe the strong disorder RG rules
consisting in the iterative elimination of the site or the link
associated with the highest frequency, and explain the relations
 with the previous approaches
\cite{hastings,imry}.
In section \ref{sec_aoki}, we compare with the Aoki RG rules 
that are exact at fixed frequency.
Our conclusions are summarized in section \ref{sec_conclusion}.

\section{ Disordered network of masses connected by springs } 

\label{models}

We consider an arbitrary network where sites are indexed by $i$.
With each site is associated a random mass $m_i$,
and with each link $<i,j>$ is associated 
a random spring constant $K_{ i,j}= K_{ j,i }$.

The scalar phonon model is defined 
by the following 
 harmonic Hamiltonian for the scalar displacements $u_{i}(t)$
\begin{eqnarray}
H = \sum_{ i} \frac{ m_{ i}}{2} \dot{u}_{ i}^2 
+ \sum_{ <i,  j>} \frac{K_{ i, j}}{2} (u_{ i}-u_{ j})^2 
\label{Hphonon}
\end{eqnarray}
The scalar assumption is very standard to simplify the analysis
\cite{Leb_spo} and means physically that vibrations along different directions
are decoupled. 
Equivalently, the model can be defined by the equations of motion 
\begin{eqnarray}
 m_{ i} \ddot{u}_{ i} = 
 \sum_{ j} K_{ i, j} 
\left(u_{ j}-u_{ i }\right)
\label{motion}
\end{eqnarray}

As stressed in \cite{hastings,imry}, this scalar phonon formulation
can be used to study various other relevant physical models :

(i) the case $m_i=1$ with random $K_{i,j}$ can be used to study the Laplacian
on the network with disordered couplings between nodes \cite{hastings,imry}.
For instance in \cite{imry}, the sites were drawn at random in a d-dimensional cube
and the $K_{i,j}$ was exponentially decaying in the Euclidean distance $r_{i,j}$.
In \cite{hastings}, the motivation was coming from the field of complex networks
(see the recent reviews \cite{complex}).

(ii) the case $m_i=1/n_i$ where $n_i$ represents the number of nodes connected to $i$ and $K_{i,j}=1$ when $(i,j)$ are neighbors on the network (with $K_{i,j}=0$
otherwise) corresponds to the usual random walk on the network.
 Here the 'disorder' comes only from the network heterogeneous structure.

\section{ Strong disorder renormalization (R.G.) approach } 

\label{sec_strong}

Strong disorder renormalization 
(see \cite{review} for a review) is a very specific type of RG
that has been first developed in the field of quantum spins :
the RG rules of Ma and Dasgupta \cite{madasgupta} 
have been put on a firm ground by D.S. Fisher 
who introduced the crucial idea of ``infinite disorder'' fixed point
where the method becomes asymptotically exact,
and who computed explicitly exact 
critical exponents and scaling functions 
for one-dimensional disordered quantum spin chains \cite{dsf}.
This method has thus generated a lot of activity for various
disordered quantum models : see the review \cite{review} for works before 2004,
as well as more recent developments concerning 
entanglement \cite{entanglement_refael,entanglement_igloi},
superfluid-insulator transition of disordered bosons \cite{superfluid},
Bose-Einstein condensation \cite{bose-einstein},
dissipation effects \cite{dissipation_gregory,dissipation_hoyos},
Hubbard model \cite{hubbard}, anyonic chains \cite{anyon}, 
fractal lattices \cite{fractal},
studies on the link with extremal statistics
\cite{extremal},
and implementation of new very efficient numerical procedures \cite{efficient}.
Strong disorder renormalization has been also
successfully applied to
various classical disordered dynamical models,
such as random walks in random media either in one dimension
\cite{sinairg,sinaibiasdirectedtraprg}, on strips
\cite{juhasz}, in two dimensions \cite{sinai2d},
or in the presence of absorbers \cite{sinaiabsorbers},
reaction-diffusion \cite{readiffrg}, 
coarsening dynamics of classical spin chains \cite{rfimrg}, 
trap models \cite{traprg}, random vibrational networks \cite{hastings,imry},
contact processes \cite{contactrg,contact_vojta},
zero range processes \cite{zerorangerg} and 
exclusion processes  \cite{exclusionrg}, non-equilibrium dynamics of 
polymers or interfaces in random media \cite{rgpsi}, statistics of valleys
in configuration space of disordered systems \cite{rgvalley}, oscillator synchronisation
\cite{phasecoupled}, and extreme value statistics of various stochastic processes
\cite{gregory}.

In this section, we describe a strong disorder renormalization procedure
for the network of masses connected by springs of Eq. \ref{Hphonon} and \ref{motion}
that generalizes the previous procedures proposed in \cite{hastings,imry}.
The principle of this RG procedure is to eliminate iteratively the highest
frequency present in the system. The elementary oscillating modes are of two
types as we now discuss.

\subsection{ Oscillating modes associated with a single spring  }

\label{single_spring}

Let us first consider 
the problem of a single spring between two masses $m_1$ and $m_2$,
when all other spring constants are supposed to be negligeable. 
The equations of motion
\begin{eqnarray}
 m_1 \ddot{u}_1 + K_{1,2} 
\left(u_1-u_2\right)&& = 0 \\
 m_2 \ddot{u}_2   
+  K_{1,2} 
\left(u_2-u_1\right) && =0
\label{motion1spring}
\end{eqnarray}
can be analyzed by introducing the center-of-mass 
displacement $u_{G}$ and the relative displacement $u_{rel}$
\begin{eqnarray}
 u_{G} && \equiv \frac{ m_1 u_1
+ m_2 u_2 }{m_1 + m_2} \\
 u_{rel} && \equiv u_2 - u_1
\label{Grel}
\end{eqnarray}
as well as their corresponding masses
\begin{eqnarray}
m_G && \equiv m_1 + m_2 \\
m_{rel} && \equiv \frac{m_1  m_2 }{m_1 + m_2}
\label{massGrel}
\end{eqnarray}
The equations of motion of Eq. \ref{motion1spring} are then decoupled
\begin{eqnarray}
 m_G \ \ddot{u}_{G} && = 0 \\
 m_{rel} \ \ddot{u}_{rel} + K_{1,2} \ u_{rel}&& = 0
\label{motion1springdecoupled}
\end{eqnarray}
One obtains the zero-mode corresponding to the global motion of the center of mass, and the oscillating model of the relative coordinate at the frequency
$\Omega_{rel}$ given by
\begin{eqnarray}
\Omega_{1,2}^2 = \frac{K_{1,2}}{m_{rel}} = K_{1,2}
 \left( \frac{1}{m_1}+ \frac{1}{m_2} \right)
\label{omegarelative}
\end{eqnarray}

If this frequency is high, this mode will not be excited by the low-energy
modes of the whole system : one may then eliminate 
 the corresponding relative mode,
i.e. we set 
\begin{eqnarray}
u_{rel}=u_2-u_1=0
\label{eliminrelative}
\end{eqnarray}
The masses $m_1$ and $m_2$ are then merged
into their center of mass $G$ of mass $m_G=m_1+m_2$ 
and of displacement $u_G=u_1=u_2$.
The equation of motion for the center of mass now takes into account
the forces of other points $i \ne (1,2)$ that were connected 
either to (1) or to (2) or to both
\begin{eqnarray}
 m_G \ddot{u}_{G} && =  \sum_{i \ne (1,2) } (K_{i,1} +K_{i,2} )
\left(u_i-u_G \right)
\label{motionGext}
\end{eqnarray}
that had been neglected in Eq. 
\ref{motion1springdecoupled} in the single spring analysis.
The equation of motion for $i \ne (1,2)$ becomes
\begin{eqnarray}
 m_i \ddot{u}_{i} && =  \sum_{n \ne (1,2) } K_{i,n} 
\left(u_n-u_i \right)
+ (K_{i,1} +K_{i,2} ) \left(u_G-u_i \right)
\label{motionjext}
\end{eqnarray}

In summary, the elimination of a spring $K_{1,2}$ associated with a high frequency 
$\Omega_{1,2}$ (see Eq. \ref{omegarelative}) consists in the merging 
of the two sites $1$ and $2$ into a single site representing their center of mass
$G$ of mass $m_G=m_1+m_2$.
The approximation of Eq. \ref{eliminrelative} yields that 
the springs constant connected to $G$ are given by
to the following renormalization rule
\begin{eqnarray}
K_{i,G}^{new} =  (K_{i,1} +K_{i,2} )
\label{rgspringrelative}
\end{eqnarray}
whereas the masses $m_i$ are unchanged.

\subsection{ Oscillating mode associated with a single mass  } 

\label{sec_singlemass}

The problem of a single mass $m_0$ in the environment of the other masses 
$m_j$ that are supposed to be fixed
( $u_i(t)=u_i$ independent of time) is of course very simple : 
from the equation of motion 
\begin{eqnarray}
 m_{0} \ddot{u}_{0}  
+ \left[ \sum_{i} K_{0,i} \right] u_{0}(t)
= \sum_{i} K_{0,i} u_{i}
\label{motion1mass}
\end{eqnarray}
one obtains that $u_{0}(t)$ will oscillate around the equilibrium value
\begin{eqnarray}
u_0^{eq}= \frac{\sum_{i} K_{0,i} u_{i}}{ \sum_{n} K_{0,n} }
\label{eq1mass}
\end{eqnarray}
 with the frequency $\Omega_0 $ given by
\begin{eqnarray}
\Omega_0^2 = \frac{1}{m_0}  \sum_{i} K_{0,i}
\label{1mass}
\end{eqnarray}

If this frequency is high, this mode will not be excited in the low-energy
modes of the whole system : one may then eliminate it. 
This amounts to say that $u_0(t)$ will adiabatically follow
the slow motion of its neighbors 
\begin{eqnarray}
u_0^{adiab}(t) \simeq \frac{\sum_{i} K_{0,i} u_{i}(t) }{ \sum_{n} K_{0,n} }
\label{adibatiq1mass}
\end{eqnarray}
The elimination of the mass $m_0$ leads to the following renormalized
of the equation of motion for the other masses $i \ne 0$
\begin{eqnarray}
 m_{i} \ddot{u}_{i} && 
= \sum_{j \ne 0} K_{i,j} (u_{j}-u_{i})
+ K_{i,0}  \left( \frac{\sum_{j} K_{0,j} u_{j}(t) }{ \sum_{n} K_{0,n} } -u_i \right) \nonumber \\
&& = \sum_{j \ne 0} \left[ K_{i,j}+ \frac{K_{i,0} K_{0,j} }{\sum_{n} K_{0,n}} \right] (u_{j}-u_{i})
\label{motionimass}
\end{eqnarray}

In summary, the elimination of a mass $m_0$ associated with a high frequency 
$\Omega_0$ (see Eq. \ref{1mass}) leads via the adiabatic approximation
of Eq. \ref{adibatiq1mass} to the following renormalizations
for the spring constants between two neighbors $(i,j)$ of the mass $m_0$
\begin{eqnarray}
K_{i,j}^{new} = K_{i,j}+ \frac{K_{i,0} K_{0,j} }{\sum_{n} K_{0,n}} 
\label{rgspring1mass}
\end{eqnarray}
whereas the masses $m_i$ are unchanged.

\subsection{ Statement of the strong disorder renormalization procedure } 

\label{fullrules}

The above analysis suggest the following strong disorder renormalization procedure :

(i) at a given RG step, there are a certain number of masses $m_i$
and a certain number of spring constants $K_{i,j}=K_{j,i}$ (with $i \ne j$).

With each spring $K_{i,j}$, one associates the frequency 
$\Omega_{i,j}$ given by
\begin{eqnarray}
\Omega_{i,j}^2 \equiv  K_{i,j} \left(\frac{1}{m_i}+ \frac{1}{m_j} \right) 
\label{omegaij}
\end{eqnarray}

With each mass $m_i$, one associates the frequency $\Omega_i$ given by
\begin{eqnarray}
\Omega_i^2 \equiv \frac{ \sum_{j} K_{i,j}}{m_i} 
\label{omegai}
\end{eqnarray}

The renormalization scale $\Omega$ is defined as the 
 highest frequency $\Omega$ remaining in the system
\begin{eqnarray}
\Omega \equiv {\rm max} \left[ \{\Omega_i , \Omega_{i,j} \} \right]
\label{rgscale}
\end{eqnarray}
among all frequencies associated with masses or spring constants.

(ii) Decimation of the mode associated with the highest frequency $\Omega$ :

(ii-a) If the highest frequency $\Omega=\Omega_{i_0,j_0}$
is associated with the spring constant $K_{i_0,j_0}$ : 
the two masses $m_{i_0}$ and $m_{j_0}$ are replaced by their center of mass $ G(i_0,j_0)$
of mass 
\begin{eqnarray}
m_{G(i_0,j_0)}= m_{i_0} + m_{j_0}
\label{Gi0j0}
\end{eqnarray}
and the spring constants linked to $i_0$ or to $j_0$ or to both are replaced by
spring constants linked to their center of mass
\begin{eqnarray}
K_{j,G(i_0,j_0)}= K_{j,i_0}+K_{j,j_0}
\label{KjGi0j0}
\end{eqnarray}

(ii-b) If the highest frequency $\Omega=\Omega_{i_0}$
is associated with the mass $i_0$ : 
the mass $m_{i_0}$ is eliminated, and the spring constants between two neighbors $(i,j)$ of $(i_0)$ are renormalized according to
\begin{eqnarray}
K_{i,j}^{new} = K_{i,j}+ \frac{K_{i,i_0} K_{i_0,j} }
{\sum_{n} K_{i_0,n}} 
\label{Kijnew}
\end{eqnarray}

(iii) Return to point (i) to update the frequencies associated with
the surviving renormalized mass and to the surviving renormalized springs.

\subsection{ Relation with the RG rules proposed by Amir, Oreg and Imry \cite{imry} }

In \cite{imry}, Amir, Oreg and Imry have proposed
 to choose at each step the largest spring constant
$K_{i,j}$ and then to apply RG rules that coincide with Eqs
\ref{Gi0j0} and \ref{KjGi0j0}.
They insist that one should choose the largest spring constant
independently of the masses, and not the highest corresponding frequency
that we introduced in Eq. \ref{omegaij}.
We do not agree on this point, since we believe that 
the frequency scale is the only well founded physical variable
to define a consistent strong disorder procedure for phonons.
We believe that for the specific
 application studied in \cite{imry},
the distribution of spring constants is so broad whereas all masses are initially equal, that the two choices are probably equivalent. 
However for other applications, we believe that the frequency choice 
is the relevant one to compare the possible excitations of various sub-systems.
 At the beginning of the RG procedure,
the considered sub-systems are made of a single mass or of a pair of masses
connected by a spring, but as the RG proceeds, the considered sub-systems
contain more and more initial masses. The frequency criterion allows
 to construct in a consistent way the appropriate renormalized modes
that will respond to an exterior low-energy excitation.

\subsection{ Relation with the RG rules proposed by Hastings \cite{hastings} } 

In \cite{hastings}, Hastings first performs the
similarity transformation \cite{dyson}
\begin{eqnarray}
 u_i = \frac{ v_i}{\sqrt{m_i }}
\label{similarity}
\end{eqnarray}
to obtain a symmetric operator from the equation of motion of Eq. \ref{motion}
\begin{eqnarray}
  \ddot{v}_{ i} = - \sum_{ j} L_{i,j} v_j
\label{motionsym }
\end{eqnarray}
with 
\begin{eqnarray}
 L_{i,i}
 \equiv  \frac{1}{m_i } \sum_j K_{ i, j}  \nonumber \\
 L_{i \ne j}
 \equiv - \frac{K_{ i, j} }{\sqrt{m_i m_j}} 
\label{lij}
\end{eqnarray}

The first RG procedure proposed in \cite{hastings} consists in choosing
the highest $L_{i,i}$ at each step : since $L_{i,i}=\Omega_i^2$ of Eq. \ref{omegai},
this is equivalent to choose the highest frequency associated with masses only
(and the frequencies 
$ \Omega_{i,j}^2 $ of Eq. \ref{omegaij} associated with links are not considered
for this choice).
The RG rule corresponding to the elimination of $L_{i_0,i_0}$ reads \cite{hastings}
(after correction of the typo concerning the sign
 of the second term in \cite{hastings})
\begin{eqnarray}
 L_{i,j}^{new} = L_{i,j} - \frac{L_{i,i_0} L_{i_0,j}}{L_{i_0,i_0}}
\label{lijrg}
\end{eqnarray}
which is equivalent to the RG rule of Eq. \ref{Kijnew}
for the spring constants
and to keeping the masses unchanged for the remaining sites.

The second RG procedure proposed in \cite{hastings} still consists in choosing
the highest $L_{i,i}$ at each step, but to treat separately its neighbor $j$
with the highest $\vert L_{i,j} \vert$ : 
the RG then consists in the diagonalization of this two-by-two submatrix 
and to project out the eigenvector of highest eigenvalue. 
In \cite{hastings}, the advantages of this procedure with respect to 
the previous rule of Eq. \ref{lijrg} were to allow the increase of the connectivity
and the variations of the masses. 
However these properties are also present in the simpler rules of Eqs
\ref{Gi0j0} and \ref{KjGi0j0} concerning the decimation of a link. 
We thus believe that the RG rules described in section \ref{fullrules}
that consider both the decimations of sites and links is physically clearer
and is able to renormalize at the same time
 spring constants, masses and connectivity
of the network in a consistent way.

\subsection{ Validity of the strong disorder renormalization procedure } 

This renormalization procedure will be consistent if the renormalization scale
$\Omega$ defined in point (i) decreases at each step, i.e. if the new
generated frequencies are smaller than the decimated frequency $\Omega$.
Moreover from what happens in other fields \cite{review},
one expects that 
the procedure will become asymptotically exact at small $\Omega$,
if the renormalized distribution of the frequencies become broader and broader
as $\Omega \to 0$.
For an arbitrary network with an arbitrary 
disorder of masses and spring constants, it is not easy to know
a priori if the procedure will remain consistent,
and if the procedure will become asymptotically exact at low frequency.
 However, this can be checked numerically for each case of interest.
It turns out that in the various cases studied numerically in \cite{hastings,imry}
the results have been found to be extremely accurate at low energies.
In the following section, we explain that this accuracy could come
from the coincidence with the zero-frequency limit of Aoki exact RG rules.

\section{ Comparison with Aoki exact renormalization procedure at fixed frequency }

 \label{sec_aoki}

For Anderson localization models, Aoki \cite{aoki80,aoki82,aokibook}
 has introduced 
 an exact real-space renormalization procedure at fixed energy which preserves the Green functions of the remaining sites. 
This procedure has been further studied for one-particle models
in \cite{lambert,us_twopoints,us_aokianderson}.
It has been also extended in configuration space
for two-particle models \cite{leadbeater} and for  manybody localization models
 \cite{us_manybody}.
This method can be extended to any eigenvalue equation, 
so it can be applied to the scalar phonon problem 
at fixed frequency as we now explain.

\subsection{ Equations of motion at fixed frequency } 

Since the equations of motion of Eq. \ref{motion} are linear, 
one may also analyse the dynamics in terms
of oscillating modes ${\hat u}_{\omega}$ of a given fixed frequency $ \omega $ :
\begin{eqnarray}
-  m_{i} \omega^2 {\hat u}_{\omega}(i) = 
 \sum_{j} K_{i,j} 
\left( {\hat u}_{\omega}(j)- {\hat u}_{\omega}(i) \right)
\label{eigen}
\end{eqnarray}

\subsection{ Aoki renormalization procedure } 

Eq. \ref{eigen} for an arbitrary site
$i=i_0$ can be used to eliminate ${\hat u}_{\omega}(i_0) $
with
\begin{eqnarray}
{\hat u}_{\omega} (i_0) = \frac{ \sum_j K_{i_0,j} {\hat u}_{\omega} (j) }
{ \sum_n K_{i_0,n} - m_{i_0} \omega^2  }
\label{eliminationi0}
\end{eqnarray}
Then the equations of the remaining sites $i \ne i_0$ 
remain of the same form as Eq. \ref{eigen}
\begin{eqnarray}
 - m_{i}^{new} \omega^2 {\hat u}_{\omega} (i) = 
\sum_{j \ne i_0} K_{i,j}^{new}(\omega) ({\hat u}_{\omega} (j)- {\hat u}_{\omega} (i)) 
\label{motionelim}
\end{eqnarray}
with the renormalized frequency-dependent springs constants
\begin{eqnarray}
 K_{i,j}^{new} (\omega)=
 K_{i,j}(\omega) + \frac{ K_{i,i_0}(\omega) K_{i_0,j}(\omega) }
{\sum_{n} K_{i_0,n}(\omega) - \omega^2 m_{i_0}(\omega) }
\label{rgK}
\end{eqnarray}
and the renormalized frequency-dependent masses
\begin{eqnarray}
  m_{i}^{new}(\omega)= m_i(\omega) + \frac{ K_{i,i_0} (\omega) m_{i_0}(\omega)  }{\sum_{n} K_{i_0,n} (\omega) - \omega^2 m_{i_0}(\omega) }
\label{rgm}
\end{eqnarray}

Let us stress the two essential differences with the strong disorder RG rules
discussed in the previous section :

(i) Aoki RG procedure is exact
 for any order in the choice of the decimated sites $i_0$,
whereas in strong disorder RG procedure it is essential to choose 
at each step the mode corresponding to the highest elementary frequency.

(ii) The price to pay is that Aoki RG concerns a fixed frequency $\omega$,
and that the renormalized parameters contain this frequency.
So for each frequency, one should restart at the very beginning the RG procedure
to compute the new renormalized couplings. 
On the contrary, the renormalized masses and springs of the strong
disorder RG procedure do not contain the frequency.

So these two RG procedures are completely different in spirit and in practice.
But it turns out that they become similar if one consider
the Aoki RG rules in the limit of zero-frequency.

\subsection{ Limit of zero-frequency of Aoki RG rules } 

In the limit of zero frequency $\omega \to 0$, the elimination of $i_0$
corresponds to the RG rule (Eq. \ref{rgK})
\begin{eqnarray}
 K_{i,j}^{new} (0)= K_{i,j}(0) + \frac{ K_{i,i_0}(0) K_{i_0,j}(0) }
{\sum_{n} K_{i_0,n}(0)  }
\label{rgKzero}
\end{eqnarray}
which coincides with the strong disorder RG rule of Eq. \ref{Kijnew}.
Note that the other RG rule of Eq. \ref{rgm} concerning the masses
disappear in the limit $\omega \to 0$ since all masses appear 
always together with $\omega^2$ in the equation of motion
(Eq. \ref{eigen}).

This coincidence between the strong disorder RG rules
and the zero-frequency Aoki exact RG rules may explain
the numerical accuracy of the strong disorder RG rules at low frequency
found in \cite{hastings,imry}.
Moreover, since Aoki procedure is exact for any order of the decimated sites,
one expects that 'bad decimations' during the strong disorder RG
('bad decimations' meaning decimations when the eliminated frequency
is not well separated from the others) 
will actually be corrected in the later stages of the renormalization 
when the neighbors will themselves be decimated.
This phenomenon has already been found in other applications of strong disorder RG,
in particular for quantum spin chains (see  Appendix E of \cite{dsf})
and for the construction of valleys in configuration space
for disordered systems (see section 3.6 of \cite{rgvalley}).

\section{ Conclusions and perspectives } 

\label{sec_conclusion}

In summary, we have proposed a strong disorder renormalization procedure
for the random phonon problem on arbitrary networks, that generalizes
 previous approaches where only site decimations \cite{hastings} or
only link decimations \cite{imry} had been considered. 
We believe that the iterative elimination of the mode of highest frequency
associated with either a mass or a spring constant 
is the appropriate formulation to study cases where both processes 
can a priori occur during the RG process
(in analogy with quantum spin chains in random transverse fields
where one decimates either sites or bonds with an energy gap criterion
\cite{dsf,review}).
To explain the accuracy of the RG rules found in the various examples
studied numerically in \cite{hastings,imry}, we have compared with
the Aoki RG rules that are exact at fixed frequency.

More generally, strong disorder RG rules 
are naturally defined on arbitrary networks,
since even if the starting point is a regular lattice of dimension $d>1$,
the RG rules changes the connectivity and thus destroys the initial
regular structure (the one-dimensional lattice is usually the only
case where the renormalized structure remains one-dimensional).
So they can be applied to disordered systems defined on complex networks.
However it turns out that complex networks constitute by themselves a disordered structure, so that even non-disordered models defined on non-regular networks
can present Griffiths phases \cite{munoz} or localization effects \cite{jmluck}.
It would be thus interesting in the future to study whether
strong disorder RG rules could be appropriate to study non-disordered systems
on complex networks that are sufficiently inhomogeneous.
It turns out that among the various RG procedures that have been recently
introduced for complex networks \cite{song,kim,radicchi,rozenfeld},
a recent proposal \cite{bizhani} consists in a purely sequential algorithm where
at each step, a node is selected at random to perform an elementary RG 
transformation : however since the nodes are not all equivalent, a natural question
is whether the selection of a node at random is really the good choice, or whether
one should choose the node or the link according to a maximal criterion,
as in strong disorder RG procedures.

\end{document}